\documentclass[preprint,showpacs,aps,prl,amsmath,amssymb]{revtex4} 
\usepackage{graphicx}
\usepackage{bm}
\begin{document} 
\title{New Constraints on the $^{18}$F($p,\alpha$)$^{15}$O Rate in Novae from the ($d,p$) Reaction}%
\author{R.\ L.\ Kozub$^{1}$, D.\ W.\ Bardayan$^{2}$, J.\ C.\ Batchelder$^{3}$, J.\ C.\ Blackmon${^2}$, C.\ R.\ Brune$^{4}$, A.\ E.\ Champagne$^{5}$, J.\ A.\ Cizewski$^{6}$, T.\ Davinson$^{7}$, U.\ Greife$^{8}$, C.\ J.\ Gross$^{2}$, C.\ C.\ Jewett$^{8}$, R.\ J.\ Livesay$^{8}$, Z.\ Ma$^{9}$, B.\ H.\ Moazen$^{1}$, C.\ D.\ Nesaraja$^{1}$, L.\ Sahin$^{5,10}$, J.\ P.\ Scott$^{1,2}$, D.\ Shapira$^{2}$, M.\ S.\ Smith$^{2}$, J.\ S.\ Thomas$^{6}$, P.\ J.\ Woods$^{7}$}
\affiliation{$^1$Department of Physics, Tennessee Technological University, Cookeville, TN 38505} 
\affiliation{$^2$Physics Division, Oak Ridge National Laboratory, Oak Ridge, TN 37831}
\affiliation{$^3$Oak Ridge Associated Universities, Bldg 6008, P.\ O.\ Box 2008, Oak Ridge, TN 37831-6374}
\affiliation{$^4$Department of Physics and Astronomy, Ohio University, Athens, OH 45701}
\affiliation{$^5$Department of Physics and Astronomy, University of North Carolina, Chapel Hill, NC 27599}
\affiliation{$^6$Department of Physics and Astronomy, Rutgers University, Piscataway, NJ 08854-8019}
\affiliation{$^7$Department of Physics and Astronomy, University of Edinburgh, Edinburgh EH9 3JZ, United Kingdom}
\affiliation{$^8$Department of Physics, Colorado School of Mines, Golden, CO 80401}
\affiliation{$^9$Department of Physics and Astronomy, University of Tennessee, Knoxville, TN 37996}
\affiliation{$^{10}$Department of Physics, Dumlupinar University, Kutahya, Turkey 43100}

\date{\today} 
%
%
\begin{abstract} 
The degree to which the ($p,\gamma$) and ($p,\alpha$) reactions destroy $^{18}$F at temperatures 1-4$\times$10$^8$ K is important for understanding the synthesis of nuclei in nova explosions and for using the long-lived radionuclide $^{18}$F, a target of $\gamma$-ray astronomy, as a diagnostic of nova mechanisms.  The reactions are dominated by low-lying proton resonances near the $^{18}$F+p threshold (E$_x$=6.411 MeV in $^{19}$Ne).  To gain further information about these resonances, we have used a radioactive $^{18}$F beam from the Holifield Radioactive Ion Beam Facility to selectively populate corresponding mirror states in $^{19}$F via the inverse $^2$H($^{18}F,p$)$^{19}$F neutron transfer reaction.  Neutron spectroscopic factors were measured for states in $^{19}$F in the excitation energy range 0-9 MeV.  Widths for corresponding proton resonances in $^{19}$Ne were calculated using a Woods-Saxon potential.  The results imply significantly lower $^{18}$F($p,\gamma$)$^{19}$Ne and $^{18}$F($p,\alpha$)$^{15}$O reaction rates than reported previously, thereby increasing the prospect of observing the 511-keV annihilation radiation associated with the decay of $^{18}$F in the ashes ejected from novae.
\end{abstract} 
\maketitle 
Understanding the synthesis of nuclei in nova explosions depends critically on knowing the rates of the $^{18}$F($p,\gamma$)$^{19}$Ne and $^{18}$F($p,\alpha$)$^{15}$O reactions at temperatures in the range 1-4$\times$10$^8 $K \cite{Coc}.  Further, owing to the relatively long half-life of the $^{18}$F radionuclide (110 min), $\gamma$ rays from electron-positron annihilation following the $\beta^+$ decay of $^{18}$F would be produced after the expanding envelope becomes transparent to the 511-keV radiation.  This should allow measurements of the abundance of $^{18}$F to be made using existing and proposed gamma ray observatories, and such measurements may provide insights about nova mechanisms.  It is therefore important to know the degree to which the ($p,\gamma$) and ($p,\alpha$) reactions destroy $^{18}$F in such explosions.  In order to deduce these reaction rates, a number of measurements using the ($p,p$) and ($p,\alpha$) reactions have been made in recent years with radioactive beams of $^{18}$F \cite{Coszach,Graulich97,Rehm,Graulich01,Bardayan00,Bardayan01,Bardayan02,Bardayan04}.  In addition, several measurements to determine properties of relevant $^{19}$Ne levels have been made with stable beams \cite{Utku,Lewis}.  There are, however, still a number of unanswered questions, especially concerning resonances at the lowest center-of-mass energies.

The levels of interest are $^{19}$Ne resonances which are just above the $^{18}$F+p threshold (6.411 MeV excitation energy). About seven levels seem to be missing in the 6.4-7.4 MeV region of excitation in $^{19}$Ne, based on what is known about the $^{19}$F mirror spectrum \cite{Utku,Butt,Tilley}.   Some of the $^{19}$Ne levels relevant to nova temperatures are inaccessible through traditional resonance scattering experiments, as the resonance energies are much below the Coulomb barrier.  It is also difficult to observe these levels via proton transfer reactions using presently existing radioactive beam facilities (in reversed kinematics).  Thus, in order to provide further insights about these $^{19}$Ne resonances, we have used the $^2$H($^{18}F,p$)$^{19}$F reaction to selectively populate neutron single-particle states in the corresponding region of the $^{19}$F mirror spectrum.  In this paper, we focus on these states of astrophysical importance.  A report on all observed states and general structure implications will be presented in a later paper.  The present work supersedes previously quoted preliminary results from this experiment \cite{Bardayan02,deSereville,Shu03}.

\begin{figure} 
\centering
\includegraphics[width=3.34in]{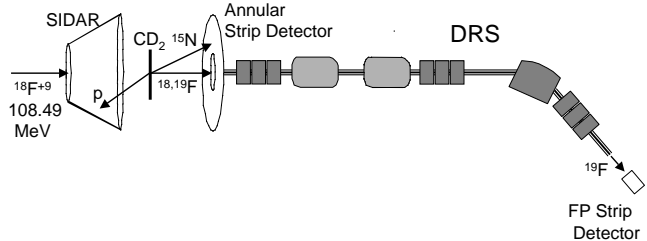} 
\caption{%
Experimental setup for $^2$H($^{18}F,p$)$^{19}$F reaction study (not to scale).} 
\label{fig:dpsetup} 
\end{figure}

A schematic diagram of the experimental setup is shown in Fig.~\ref{fig:dpsetup}.  A 160(10) $\mu$g/cm$^2$ (CD$_2)_n$ target of 98\% enrichment was bombarded for $\sim$3 days with an isotopically pure, 108.49 MeV $^{18}$F$^{+9}$ beam at an intensity of $\sim$5$\times$10$^5$/s.  The beam was produced at the ORNL Holifield Radioactive Ion Beam Facility (HRIBF) as described in Ref. \cite{Bardayan01}.  Using a silicon strip detector array (SIDAR) \cite{Bardayan01} of $\sim$500 $\mu$m thickness, light charged particles were detected in the laboratory angular range of $118^{\circ} - 157^{\circ}$, corresponding to ``forward'' center-of-mass angles in the range $8^{\circ} - 27^{\circ}$.  The beam energy was selected to be high enough for direct reaction models, yet low enough to allow all the protons to be stopped in the SIDAR.  A silicon strip detector at the focal plane of the Daresbury Recoil Separator (DRS) \cite{Fitzgerald} was used to detect particle-stable recoils having A=19 in coincidence with the SIDAR.  This coincidence efficiency was essentially 100\% of the +9 charge state fraction for recoil angles $<1.6^{\circ}$ and $>$70\% overall for particle-stable final states.  Other recoils, from higher, $\alpha$-decaying states in $^{19}$F, were detected in coincidence just downstream from the target with an annular strip detector.  This detector was also used for data normalization.  Beam current normalization was achieved by directly counting beam particles at low intensity with a retractable silicon surface barrier detector placed temporarily at $0^{\circ}$.  The overall uncertainty in normalization, estimated to be $\sim$10\%, is owing mostly to uncertainty in target thickness.  Independent internal energy calibrations were obtained for each laboratory angle by using excitation energies of the well-known levels at 1.554038(9), 4.377700(42), and 5.1066(9) MeV in $^{19}$F \cite{Tilley}.  This allowed excitation energies in the region of interest to be determined with uncertainties $\sim$10 keV.
%
%

\begin{figure} 
\centering
\includegraphics[width=2.50in]{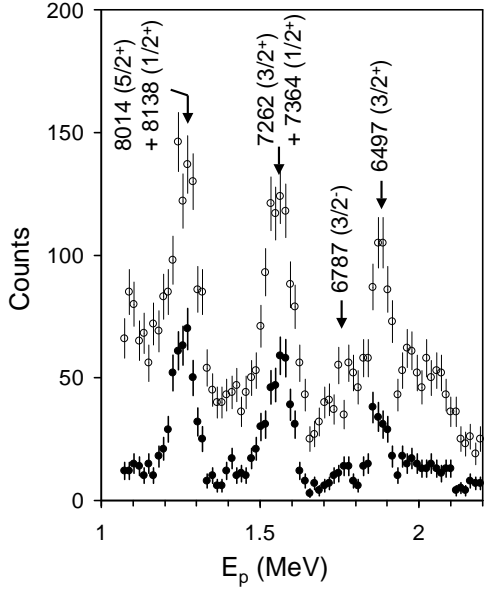}
\caption{%
Spectra from the 6 SIDAR strips corresponding to $147^{\circ}$ lab angle as a function of proton energy, displayed in 15-keV bins.  Open points are singles data and solid points are events in coincidence with the annular strip detector.  Excitation energies (in keV), spins, and parities are taken from Ref. \cite{Tilley}.}
\label{fig:spectra} 
\end{figure}

Singles and coincidence spectra in the ($\alpha$-decaying) region of importance for novae are shown for a laboratory angle of $147^{\circ}$ in Fig.~\ref{fig:spectra}.  The coincidence spectra contain three main groups at about 6.5, 7.3, and 8.1 MeV excitation energy.  The coincidence efficiencies are roughly 50\%, 60\%, and 70\%, respectively.  Our internal energy calibrations allow identifications of these states which are consistent with known levels in $^{19}$F, as indicated in Fig.~\ref{fig:spectra}.  
%
%

\begin{figure}
\centering
\includegraphics[width=3.00in]{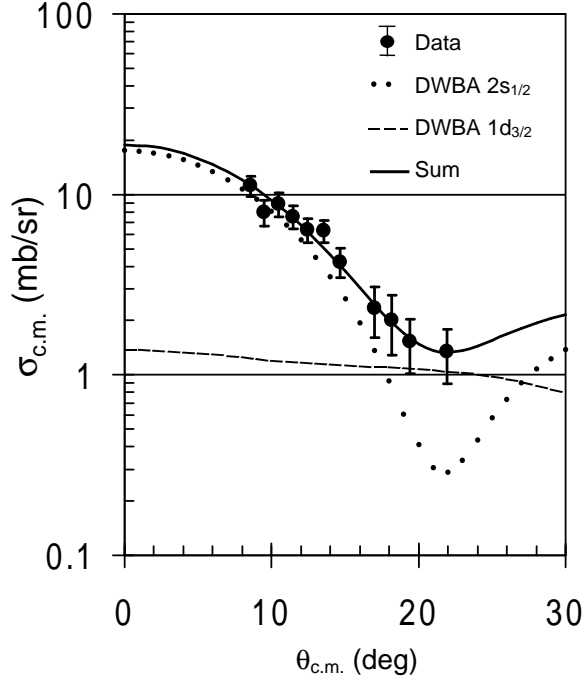}
\caption{%
Angular distribution for ($d,p$) transfer to the 6497-keV, 3/2$^+$ state in $^{19}$F.  Curves are DWBA calculations using parameters from Ref. \cite{Barrows}.}
\label{fig:6497} 
\end{figure}

Angular distributions have been extracted for all observed proton groups, and distorted-wave Born approximation (DWBA) calculations have been performed with the code DWUCK5 \cite{Kunz} using optical model and bound state parameters from Barrows {\em et al.} \cite{Barrows}, who studied the $^{19}$F($d,n$)$^{20}$Ne reaction at several bombarding energies.  In Ref. \cite{Barrows}, searches were done on the optical model and bound state parameters to provide a best fit to the actual ($d,n$) reaction data.  This exercise produced good fits over a wide angular range for $\ell$=0 and $\ell$=2 transitions, and the spectroscopic factors were quite reasonable.  We have performed DWBA calculations for our ($d,p$) data using the parameters of Ref. \cite{Barrows} for their highest bombarding energy (6.065 MeV).  This resulted in good fits to our data over the entire range of excitation energies and in reasonable spectroscopic factors.  Since J$^{\pi}$=1$^+$ for the $^{18}$F ground state, more than one angular momentum transfer is allowed for a given final state.  Indeed, with the DWBA parameters of Ref. \cite{Barrows}, including both $\ell$=0 and $\ell$=2 components improves the fits at larger angles in the distributions for 3/2$^+$ and 1/2$^+$ states, although the $\ell$=2 curves are rather featureless over the angular range measured.  The angular distribution for the 6497-keV 3/2$^+$ level is shown in Fig.~\ref{fig:6497} along with DWBA calculations for 2s$_{1/2}$ and 1d$_{3/2}$ transfers.  Spectroscopic factors for this and other observed states of astrophysical interest are summarized in Table~\ref{tab:table1}.  Uncertainties in S$_n$ shown for observed states are due to curve fitting and data normalization only.  Listings for some unobserved states are given 100\% errors to reflect upper limits.
%
%
%
%
%
\begin{table*}
\caption{\label{tab:table1}Summary of data for presumed mirror states in $^{19}$Ne and $^{19}$F.  $^{19}$Ne levels with tentative mirror assignments are shown in brackets ().  Spectroscopic factors for $^{19}$F and proton widths deduced from them (assuming S$_p=$S$_n$) are from the present work.  Except as noted, $^{19}$F excitation energies and J$^{\pi}$ assignments are from Ref. \cite{Tilley}.  Other values are from Bardayan {\em et al.}\cite{Bardayan02}}
\begin{ruledtabular}
\begin{tabular}{cccccccccc}
 &\multicolumn{3}{c}{$E_x$ (keV)}&&\multicolumn{3}{c}{$S_n$ ($^{19}$F)}&\multicolumn{2}{c}{$^{19}$Ne}\\
 $E_r$ (keV)&$^{19}$Ne&$^{19}$F&d($^{18}F,p$)\footnotemark[1] &J$^{\pi}$&1$p_{1/2}$&2$s_{1/2}$&1$d_{3/2}$&$\Gamma_p$ (keV)&$\Gamma_\alpha$ (keV)\\ \hline
8&(6419)&6497&6497(10)&3/2$^+$& &0.12(2)&0.13(6)&2.2(4)$\times$10$^{-37}$&0.5(5)\\
26&6437&6429& &1/2$^-$& & & &1.1$\times$10$^{-20}$&220(20)\\
38&(6449)&6528& &3/2$^+$& &0.03(3)& &4(4)$\times$10$^{-15}$&4(4)\\
287&6698&6838& &5/2$^+$& & &0.01(1)&1.2(12)$\times$10$^{-5}$&1.2(10)\\
330&6741&6787&6795(15)&3/2$^-$&0.05(1)& & &2.22(69)$\times$10$^{-3}$\footnotemark[2]&2.7(23)\\
665&7076&7262\footnotemark[3]& &3/2$^+$& &0.16(3)&0.05(5)&15.2(10) \footnotemark[2]&24(2)\\
 & & &7306(10)\\
 & &7364\footnotemark[3]& &1/2$^+$& &0.16(3)&0.05(5)\\
 & &8014\footnotemark[3]& &5/2$^+$& & &0.19(4)\\
 & &8138\footnotemark[3]& &1/2$^+$& &0.32(6)\\
\end{tabular}
\end{ruledtabular}
\footnotetext[1]{Present work.}
\footnotetext[2]{See Ref. \cite{Bardayan02}.} 
\footnotetext[3]{Member of an unresolved doublet in ($d,p$) data.}
\end{table*}

It has been widely assumed \cite{Coc,Bardayan02,Utku,deSereville,Shu03} that the 6497- and 6528-keV 3/2$^+$ states are the isospin mirrors of the 8- and 38-keV proton resonances (6419 and 6449 keV excitation energies) in $^{19}$Ne, respectively, although the spins and parities of the $^{19}$Ne states have not been determined experimentally.  Nevertheless, working under this assumption, it is clear that the 38-keV resonance in $^{19}$Ne would dominate the reaction rate for T$<$0.27 GK if that state has a sizeable share of the $\ell$=0 single particle strength \cite{Bardayan02}. On the other hand, there would be almost no contribution to the rate if the $\ell$=0 strength is concentrated in the 8-keV resonance.  Our measured excitation energy for the $^{19}$F group is 6497(10) keV and the peak width seems to match our experimental resolution [$\sim$50 keV (lab) at $147^{\circ}$; see Fig.~\ref{fig:spectra}], so there is no evidence for the 6528-keV level in our data. (A limit of S$<$0.06 is extracted using our centroid uncertainty.)  This is contrary to the result of de S\'{e}r\'{e}ville {\em et al.} \cite{deSereville}.  In a $^2$H($^{18}F,p$)$^{19}$F study at a much lower beam energy (14 MeV) than the present work (108 MeV), they obtained S$_n$=0.21 for the 6.5-MeV group and indicated some preference for the excitation of the 6528-keV level.  Further, in a recent reanalysis of these low-energy data \cite{deSereville04}, it was concluded that the dominant contribution is from the 6528-keV level.  However, their data contain strong multi-step excitations, such as the 9/2$^+$ state at 2.78 MeV \cite{deSereville04}, which are observed to be very weak in our spectra.  Indeed, the yield of the 2.78-MeV state is used in \cite{deSereville04} to estimate the compound nucleus contribution to the 6.5-MeV group, which is then used to reduce their spectroscopic factor to 0.17.  It seems clear that the reaction mechanisms involved are very different at the two bombarding energies.

Since we see no evidence for the 6528-keV component, our result indicates that the $\ell$=0 strength near the proton threshold in $^{19}$Ne may well be concentrated in a proton-bound state, or perhaps in the 8-keV resonance.  Using the prescription outlined in Langanke {\em et al.}\cite{Langanke}, Shu \cite{Shu} has estimated a Thomas-Ehrman shift of $\sim$160 keV for $\ell$=0 resonances having S$\sim$0.1.  Although this calculation does not account for details of the state's parentage, the result is roughly consistent with the shift of 186 keV observed for the 7262-7076-keV $^{19}$F-$^{19}$Ne mirror pair (see below).  If the 38-keV $^{19}$Ne resonance is the mirror of the 6497-keV $^{19}$F state, the shift would be only 48 keV.  We therefore tentatively assign the analog to be the 8-keV resonance in $^{19}$Ne (78-keV shift), although the 38-keV resonance cannot be ruled out as a possible mirror.  Further, we observe no evidence for either the 5/2$^+$ state at 6838 keV excitation (S$<$0.02) or a broad 1/2$^-$ level at either 6429 keV \cite{Tilley} or 6536 keV \cite{Bardayan05}, the mirror analogs of which could also be of importance to nova reaction rates.
%
\begin{figure}
\centering 
\includegraphics[width=3.34in]{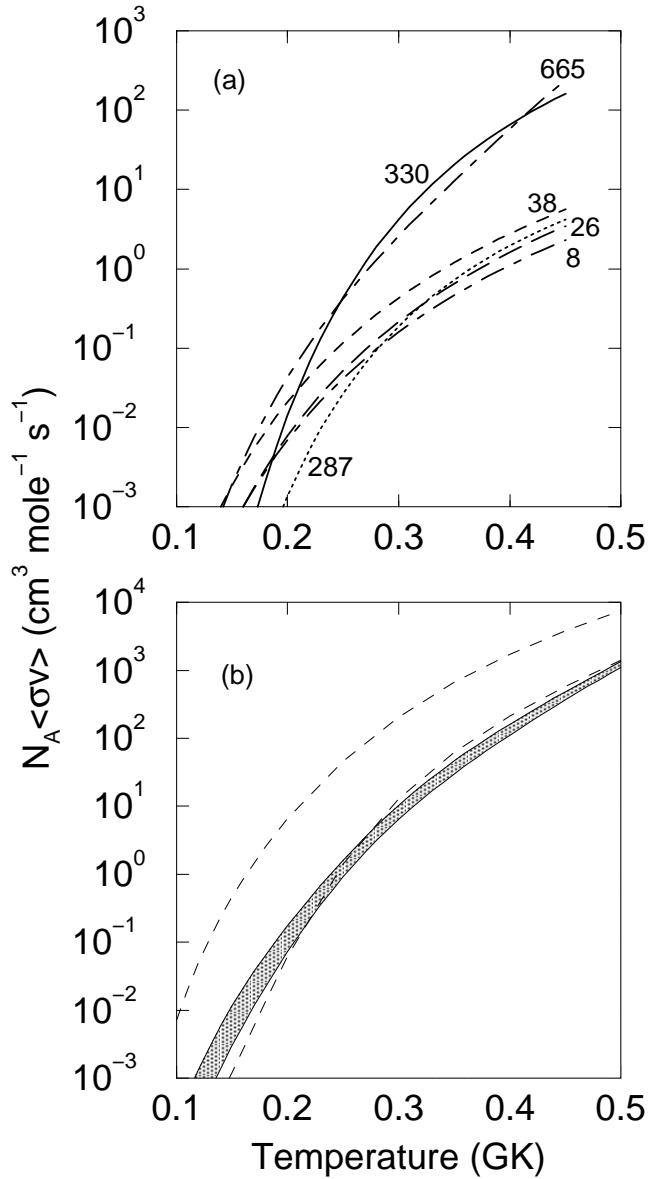}
\caption{%
(a)The astrophysical $^{18}$F($p,\alpha$)$^{15}$O reaction rate at nova temperatures labeled with the energies (in keV) of the contributing resonances, using widths and mirror assignments from Table~\ref{tab:table1}. (b) The total $^{18}$F($p,\alpha$)$^{15}$O reaction rate using results from the present work is shown as a shaded band, which includes the rates shown in (a) and those with the mirror assignments of the 8- and 38-keV resonances reversed (see text).  The dashed lines show limits from Ref. \cite{Coc}.}
\label{fig:rates} 
\end{figure} 

The 3/2$^+$ level at 7076 keV excitation in $^{19}$Ne has a relatively large proton decay branch ($\Gamma_p$/$\Gamma$=0.39(2) \cite{Bardayan01}), and Fortune and Sherr  \cite{Fortune} have calculated a spectroscopic factor of $\sim$0.5 \footnote{Using the bound state parameters of Ref. \cite{Barrows} and the resonance energy of Ref. \cite{Bardayan01}, we calculate S$_p$=0.36.}, so it should have strong $\ell$=0 analog in $^{19}$F.  The mirror state was presumed to be the state observed at 7.101 MeV in a $^{15}$N($\alpha,\gamma$)$^{19}$F experiment \cite{Butt}.  However, we see no evidence for the 7.101 MeV state in our ($d,p$) data.  Fortune and Sherr \cite{Fortune} predicted the state would lie higher in energy, at 7.4(1) MeV, owing to the relatively lower Coulomb energy associated with an extended $\ell$=0 proton wave function in $^{19}$Ne.  In our data, the 7.3-MeV group appears to be composed of known (but unresolved) states at 7262 (3/2$^+$) and 7364 keV (1/2$^+$) in $^{19}$F (Fig.~\ref{fig:spectra}).  The angular distribution of this group is dominated by $\ell$=0 components, and a two-peak fitting analysis indicates that the angular distributions of the two states are very similar in shape with a yield ratio of $\sim$2:1, suggesting roughly equal spectroscopic factors of 0.16(3) (Table~\ref{tab:table1}).  Thus, although its spectroscopic factor is significantly less than 0.5, it would appear that the 7262-keV level in $^{19}$F is a likely candidate for the analog of the 7076-keV level, roughly in agreement with Ref. \cite{Fortune}, and we make this assignment in Table~\ref{tab:table1}.  However, isospin symmetric reactions with stable beams have indicated an analog connection between the 7262-keV level and the 7238-keV level in $^{19}$Ne \cite{Utku,Lewis}, so it would appear there may be another, as yet undiscovered 3/2$^+$ state in $^{19}$F near this excitation energy \footnote{If we treat the 7.3-MeV group as a single, broad 3/2$^+$ state in our ($d,p$) data, its energy would be 7306(10) keV and the 2s$_{1/2}$ spectroscopic factor would be 0.24(5).}.  Further, the significant $\ell$=0 strength observed in the 7364- and 8138-keV states in $^{19}$F in the present work would suggest that there may be more strong $\ell$=0 resonances in the corresponding region of $^{19}$Ne, and these could make important contributions to the $^{18}$F+p reaction rate at higher temperatures (T$>$1.0 GK).  There is evidence for one such resonance in a recent thick-target study of the $^{18}$F($p,p$)$^{18}$F excitation function \cite{Bardayan04}.  
  
We have calculated the $^{18}$F($p,\alpha$)$^{15}$O reaction rates for temperatures in the range 0.1-0.4 GK in the same manner as Ref. \cite{Bardayan02}, using spectroscopic factors and uncertainties as listed in Table~\ref{tab:table1} for resonances in $^{19}$Ne for which the proton widths have not been measured directly (see Fig.~\ref{fig:rates}).  We have not included the $^{18}$F($p,\gamma$)$^{19}$Ne rates, as they are believed to be much slower than the $^{18}$F($p,\alpha$)$^{15}$O rate in novae \cite{Bardayan01,Utku}.  In the present work, we have calculated the proton single particle widths using a Woods-Saxon well having the same radius and diffuseness parameters as were used for the neutron bound state in the DWBA calculations.  This procedure was first suggested by Schiffer \cite{Schiffer} and has been used successfully in subsequent works (see, e.g., Refs. \cite{Kozub,Hale}).  An advantage of this technique is the fact that the calculated proton width, $\Gamma_p=S_n\Gamma_{sp}$, is relatively independent of potential parameters, provided the same parameters are used to calculate S$_n$ and $\Gamma_{sp}$.  This change in $\Gamma_{sp}$ alone accounts for a factor of $\sim$2 reduction in rates at the lowest temperatures from earlier work \cite{Coc,Bardayan02,deSereville}, where Wigner-Teichman limits \cite{Teichman} were used for single particle widths.  

As was the case in Refs. \cite{Coc,Bardayan02}, interference effects among $3/2^+$ states were not included in our rate calculations.  Among the four possibilities for relative phases of the interference terms, one would cause about a $\times$2 increase in the reaction rate in the 0.1-0.2 GK range, one would cause essentially no change, and the other two would cause a decrease of $\times$3 to $\times$5.  A factor of relevance to this discussion is the width of the 6497-keV level in $^{19}$F.  Fortune \cite{Fortune03} has extracted a width of 23(4) keV for this state via analysis of $^{15}$N($^7Li,t$)$^{19}$F data, but the resonance was not seen in the $^{15}$N($\alpha,\alpha$)$^{15}$N experiment of Smotrich, {\em et al.}\cite{Smotrich}, even though they were sensitive to resonances having widths $\gtrsim$1 keV.  We are therefore led to suspect that there may be some strong collective effects in the ($^7Li,t$) data analyzed by Ref. \cite{Fortune03}.  For this reason, we are using 0.5(5) keV for the width (to reflect an upper limit of 1 keV) in our calculations.  Also, resonances above the 665-keV resonance were not included in the calculations, but they are not expected to have significant impact at these lower temperatures.  

Upper and lower rates were calculated by varying the contribution from each resonance within its tabulated uncertainty (Table~\ref{tab:table1}) and then combining the resulting rate variations in quadrature.  Since the possibility that the 38-keV resonance is the mirror to the 6497-keV state in $^{19}$F cannot be excluded, a similar calculation was done with the mirror assignments of the 8- and 38-keV resonances reversed.  The combined rate band, which includes both possibilities, is plotted in Fig.~\ref{fig:rates}(b).  The nominal total rate is reduced from that of Ref. \cite{Bardayan02} by a factor of 1.5 or 4 in the 0.1-0.2 GK range, depending on whether the $\ell$=0 strength is assigned to the 38- or 8-keV resonance, respectively.  The rate is about 2-5 times smaller than that of Coc {\em et al.} \cite{Coc} in the same temperature range. Recent nucleosynthesis calculations by Smith {\em et al.} \cite{Smith} using these new rates show a factor of $\sim$1.6 increase in the $^{18}$F mass fraction following novae if the $\ell$=0 strength lies in the 8-keV resonance.  This would allow observatories to view novae at larger distances, and thus over a significantly larger ($\sim$2$\times$) spatial volume than estimated previously.  However, the true location in $^{19}$Ne of the mirror of the 6497-keV 3/2$^+$ state is obviously a critical issue, one that can be settled only with proton transfer data taken at a bombarding energy which is high enough for direct reaction models to be viable.  

In summary, using the $^2$H($^{18}F,p$)$^{19}$F reaction, we have measured neutron spectroscopic factors (or upper limits) for the mirror analogs of eight levels of potential astrophysical importance in $^{19}$Ne.  With reasonable mirror assignments, our measurements and updated width calculations show that the destruction of $^{18}$F via proton-induced reactions in novae is considerably less than recent predictions.  This is favorable news for the prospect of observing the characteristic 511-keV annihilation radiation associated with the decay of $^{18}$F in the ejected ashes of such events.

Thanks are owed to the staff of the HRIBF for making this experiment possible and to P. D. Parker and D. J. Millener for helpful discussions.  This work is supported in part by U. S. DOE Grant Nos.\ DE-FG02-96ER40955 (TTU) and DE-FG02-93ER40789 (CSM) and by the National Science Foundation (RU).  ORNL is managed by UT-Battelle, LLC, for the U. S. DOE under Contract No.\ DE-AC05-00OR22725.

%

\end{document}